\documentstyle[prb,aps]{revtex}
\begin{document}
\title{A Statistical Mechanical Approximation for the 
Calculation of Time Auto-Correlation Functions}
\author{Jeppe C. Dyre}
\address{Department of Mathematics and Physics (IMFUFA), 
Roskilde University, POBox 260, DK-4000 Roskilde, DENMARK}
\date{\today}
\maketitle{}
\begin{abstract}
This paper \cite{note} considers the problem of estimating the
time auto-correlation function for a quantity that is defined in
configuration space, given a knowledge of the mean-square
displacement as function of time in configuration space.
The problem is particularly relevant for viscous liquids,
where the interesting time-scales are often beyond those
reachable by computer simulation.
An approximate formula is derived which reduces the calculation
of the time auto-correlation function to a "double canonical"
average.
In this approximation, in the case of Langevin dynamics the
mean-square displacement itself may be evaluated from the "double
partition function" .
The scheme developed is illustrated by computer simulations of
a simple one-dimensional system at different temperatures,
showing good agreement between the exact time auto-correlation
functions and those found by the approximation.
\end{abstract}
\pacs{}
The calculation of a time auto-correlation function
\cite{zwa65,boo80,han86} is a straightforward matter in any
computer simulation tracing the time evolution of a system
\cite{all87,cic87,all93}.
However, computer simulations are not feasible today on time
scales longer than microseconds.
These time scales are relevant, e.g., for viscous liquids
approaching the glass transition.
Therefore, one cannot by simulation calculate a number of
experimentally easily accessible quantities in highly viscous
liquids like the frequency-dependent viscosity \cite{bir87}, bulk
modulus \cite{chr94}, dielectric constant \cite{wil75}, or
specific heat \cite{chr85,bir85} (quantities that all via the
fluctuation-dissipation theorem \cite{zwa65,boo80,han86} are
given as Laplace transforms of an equilibrium time
auto-correlation function.
In this situation one would like to have an approximate theory
at hand.
Focussing only on time auto-correlation functions of
quantities $A(X)$ that are defined in configuration space,
$X=(X_1,...,X_n)$, an approximation is proposed below, based on
an ansatz for the joint probability of initial point $X$ at
$t=0$ and final point $X'$ after time $t$, denoted by
$P(X,X';t)$.
This joint probability directly determines the time
auto-correlation function via

\begin{equation}\label{1}
\left<A(0)A(t)\right>\ =\ 
\int dX dX' A(X)A(X') P(X,X';t)\,.
\end{equation}

If $Z$ is the configurational partition function, given in
terms of $\beta=1/(k_BT)$ and the potential energy $U(X)$ as

\begin{equation}\label{2}
Z(\beta)\ =\ 
\int dX e^{-\beta U(X)}\,,
\end{equation}
and $G(X\rightarrow X';t)$ is the Green's function, the exact
expression for the joint probability is

\begin{equation}\label{3}
P(X,X';t)\ =\ 
\frac{e^{-\beta U(X)}}{Z}G(X\rightarrow X';t)\,.
\end{equation}
However, Eq.\ (\ref{3}) is not very useful unless the Green's
function is known.  
Note that the principle of detailed balance - expressing the
time-reversal invariance of the equations of motion - implies
that $P(X,X';t)\ =\ P(X',X;t)$, a requirement that any
approximation should also satisfy.

The exact method for calculating the time auto-correlation
function $\left<A(0)A(t)\right>$ is illustrated in Fig. 1
showing the path in configurational space.
At a number of times $t_1,...,t_p$ one computes the product
$A(X(t_j))A(X(t_j+t))$, and $\left<A(0)A(t)\right>$ is the
average of this quantity as $p\rightarrow\infty$.
Assuming here and henceforth that the $X_i$'s are simple
rectangular coordinates and that $\left<A\right>=0$, one
always expects $\left<A(0)A(t)\right>\rightarrow 0$ as
$t\rightarrow\infty$.
This loss of correlation after long time comes about because
the point in configuration space at time $t$, $X'$ , for large
$t$ is far away from the initial point $X$.
A measure of the distance travelled in time $t$ is provided by
the mean-square displacement, $\left<\Delta X^2(t)\right>$.
This quantity, of course, is defined by 

\begin{equation}\label{4}
\left<\Delta X^2(t)\right>\ =\ 
\left<(X-X')^2\right>\equiv
\sum_{i=1}^n \left<(X_i-X'_i)^2\right>\,.
\end{equation}

Assuming that the mean-square displacement itself is a known
function of time, the idea is now to estimate
$\left<A(0)A(t)\right>$ via the "spatial" auto-correlation
of $A$ in configuration space evaluated at a distance equal to
$\sqrt{\left<\Delta X^2(t)\right>}$.
Before proceeding, we briefly discuss the physics of this way
of thinking about the time auto-correlation function.
A simple case is when the mean-square displacement is
proportional to time (for large times this is, of course,
always the case).
In this case, if the "spatial" correlation of $A$ has a
Gaussian distance decay, the time auto-correlation function is
a simple exponential, corresponding to Debye relaxation.
If, however, the spatial correlation of $A$ has an exponential
decay, the time auto-correlation function is a stretched
exponential with exponent $1/2$.
The latter case gives a reasonable fit to many experiments on
viscous liquids \cite{boh93}.
The above picture of decomposing the time-autocorrelation
function into A) a "geometric" correlation and B) the distance
travelled in a given time, is in harmony with another
well-known property of viscous liquids. 
In these systems all linear relaxation functions determine
roughly the same average relaxation time, a time which increases
dramatically upon cooling.
In the above "geometric" picture this is simply a consequence
of the motion slowing down in configuration space, whereas
"spatial" correlations in many cases probably only change little
upon cooling.

We now turn to the problem of estimating the joint probability
$P(X,X';t)$.
In the thermodynamic limit ($n\rightarrow\infty$) the relative
fluctuations in the mean-square displacement go to zero, and
therefore the distance between $X=X(t_j)$ and $X'=X(t_j+t)$ is
"precisely" equal to $\sqrt{\left<\Delta X^2(t)\right>}$.
Similarly, the relative fluctuations in potential energy go to
zero, so the potential energy of both points $X$ and $X'$ is
"equal to" $\left<U\right>=-\frac{\partial\ln
Z}{\partial\beta}$.
Now, the basic idea is an ansatz for $P(X,X';t)$ that assumes
equal probability for all pairs of points with correct distance
and correct potential energy.
Thus, 

\begin{equation}\label{5}
P(X,X';t)\propto
\delta[(X-X')^2-\left<\Delta X^2(t)\right>]\ 
\delta[U(X)-\left<U\right>]\ 
\delta[U(X')-\left<U\right>]\,.
\end{equation}
In the thermodynamic limit there is "equipartition" between
$U(X)$ and $U(X')$, and the last two delta functions may be
replaced by a single delta function, leading to

\begin{equation}\label{6}
P(X,X';t)\propto
\delta[(X-X')^2-\left<\Delta X^2(t)\right>]\ 
\delta[U(X)+U(X')-2\left<U\right>]\,.
\end{equation}
The next step is to convert Eq.\ (\ref{6}) to a "canonical"
form, which is computationally much more convenient than the
present "microcanonical" form.
This is done by replacing the first delta function by
$\exp[-a(X-X')^2]$ where $a$ is a Lagrangian multiplier
adjusted to give the correct mean-square displacement.
Similarly, the second delta function is replaced by
$\exp[-b[U(X)+U(X')]]$ where $b$ is adjusted to ensure that
the average of $U(X)+U(X')$ is $2\left<U\right>$.
If the following "double partition function"

\begin{equation}\label{7}
D(a,b)\ =\ 
\int dX dX' e^{-a(X-X')^2-b[U(X)+U(X')]}
\end{equation}
is introduced, the final expression for approximately calculating
the time auto-correlation function is

\begin{equation}\label{8}
\left<A(0)A(t)\right>\ =\ 
\int \frac{dX dX'}{D(a,b)}A(X)A(X')
e^{-a(X-X')^2-b[U(X)+U(X')]}\,.
\end{equation}
In the thermodynamic limit Eq.\ (\ref{8}) is equivalent to
calculating the average of Eq.\ (\ref{1}) by using Eq.\
(\ref{5}).

The two parameters $a$ and $b$ are determined in the following
way.
First, $b=b(a)$ is found from the condition that the average
joint potential energy of initial and final point is 
$2\left<U\right>$.
Thus, $b=b(a)$ is determined from the condition that this
average is independent of $a$:

\begin{equation}\label{9}
\frac{d}{da}
\frac{\partial\ln D}{\partial b}\ =\ 0\,.
\end{equation}
Since $\frac{d}{da}=\partial_a+\frac{db}{da}\partial_b$
[with standard abbreviation for partial derivatives],
the expansion of Eq.\ (\ref{9}) leads to the following first
order differential equation for $b=b(a)$:

\begin{equation}\label{10}
\frac{db}{da}\ =\ 
\frac{\partial_a D\partial_b D - D \partial_{ab}D}
{D\partial_b^2 D-(\partial_b D)^2}\,.
\end{equation}
Once the function $b(a)$ has been determined, $a=a(t)$ is
found from requiring that the mean-square displacement
calculated from $D(a,b(a))$ is correct:

\begin{equation}\label{11}
-\frac{\partial\ln D}{\partial a}=\ 
\left<\Delta X^2(t)\right>\,.
\end{equation}
The short and long time limits are determined as follows.
For $a(t)$ one clearly has

\begin{equation}\label{12}
a(t=0)=\infty\ {\rm and}\ 
a(\infty)=0.
\end{equation}

In the limit of large times $X$ and $X'$ are far apart and
$U(X)$ is uncorrelated with $U(X')$.
In this limit $b=\beta$:

\begin{equation}\label{13}
b(a=0)\ =\ \beta\,.
\end{equation}
In the short time limit the points $X$ and $X'$ are close.
Thus, $P(X,X';t)\propto\delta[X-X']\exp[-2bU(X)]$ for
$t\rightarrow 0$ and Eq.\ (\ref{1}) yields

\begin{equation}\label{14}
\lim_{t\rightarrow 0} \left<A(0)A(t)\right>\ =\ 
\frac{\int dX A^2(X) e^{-2bU(X)}}{\int dX e^{-2bU(X)}}\,.
\end{equation}
For this to give the correct canonical average one
must have $b=\beta/2$, i.e.,

\begin{equation}\label{15}
b(a=\infty)\ =\ \frac{\beta}{2}\,.
\end{equation}

As we will show now, the short time behavior of $a(t)$ may be
derived directly from the equations of motion.
In the case of Newtonian dynamics, the Green's function at
short times is easily found from the integrated equations of
motion, where the momentum is distributed according to a
Gaussian (for simplicity all particles are assumed to have
equal mass, $m$)

\begin{equation}\label{16}
G(X\rightarrow X';t)\ \propto\ 
\exp\left[-\frac{\beta}{2m}\sum_{i=1}^n
\left(\frac{m}{t}(X'_i-X_i)+\frac{1}{2}\partial_i U\ t
\right)^2\right]\,.
\end{equation}
To first order in $t$ this becomes

\begin{equation}\label{17}
G(X\rightarrow X';t)\ \propto\ 
\exp\left[-a(t)(X-X')^2
-\frac{\beta}{2} [U(X')-U(X)]\right]\,,
\end{equation}
where

\begin{equation}\label{18}
a(t)\ =\ 
\frac{\beta m}{2t^2}\ \,[{\rm Newtonian\ dynamics,}\ 
t\rightarrow 0]\,.
\end{equation}
Note that via Eq.\ (\ref{3}) this Green's function confirms
the form of Eq.\ (\ref{8}) for $t\rightarrow 0$,
as well as the boundary condition Eq.\ (\ref{15}).
Next, we consider the case of Langevin dynamics,

\begin{equation}\label{19}
\dot{X_i}\ =\ -
\mu \frac{\partial U}{\partial X_i} + \xi_i(t)\,,
\end{equation}
with the standard Gaussian white noise term \cite{van81}
$\left<\xi_i(t)\xi_j(t')\right>=2\mu k_BT 
\delta_{i,j}\delta(t-t')$.
From the equations of motion one finds that, because the
integrated noise term is Gaussianly distributed,

\begin{equation}\label{20}
G(X\rightarrow X';t)\ \propto\ 
\exp\left[-\frac{\beta}{4\mu t}\sum_{i=1}^n
\left(X'_i-X_i+\mu\ \partial_i U\ t\right)^2\right]\,.
\end{equation}
At short times this again leads to Eq.\ (\ref{17}),  
where however now

\begin{equation}\label{21}
a(t)\ =\ 
\frac{\beta}{4 \mu t},\ 
[{\rm Langevin\ dynamics,}\ 
t\rightarrow 0]\,.
\end{equation}

In the case of Langevin dynamics Eq.\ (\ref{8}) may be applied
to the calculation of the force-force time auto-correlation
function.
This leads to an equation that in principle allows a
calculation of ${\left<\Delta X^2(t)\right>}$ directly
form the double partition function:
The mean-square displacement in time $t$ is given by
(sum over repeated indices)

\begin{equation}\label{22}
{\left<\Delta X^2(t)\right>}\ =\ 
\int_0^t dt'\int_0^t dt''
\left<\dot X _i(t')\dot X _i(t'')\right>\,.
\end{equation}
Since the noise terms are uncorrelated at different times, 
Eqs.\ (\ref{19}) and (\ref{22}) imply

\begin{equation}\label{23}
\frac{d^2}{dt^2}{\left<\Delta X^2(t)\right>}\ \ =\ 
2\left<\dot X _i(0)\dot X _i(t)\right>\ =\ 
2\mu^2 \left<\partial_iU(0)\partial_iU(t)\right>\,.
\end{equation}
From Eq.\ (\ref{8}) the force-force time auto-correlation
function is rewritten as

\begin{equation}\label{24}
\left<\partial_iU(0)\partial_iU(t)\right>\ =\ 
\frac{1}{b^2}\int \frac{dXdX'}{D(a,b)}
\left[\partial_ie^{-bU(X)}\right]
\left[\partial '_ie^{-bU(X')}\right]
e^{-a(X-X')^2}\,.
\end{equation}
By partial integrations one finds

\begin{equation}\label{25}
\left<\partial_iU(0)\partial_iU(t)\right>\ =\ 
-4\ \frac{a^2}{b^2}\ \int \frac{dXdX'}{D(a,b)}\ 
(X-X')^2\ 
e^{-a(X-X')^2-b[U(X)+U(X')]}\ =\ 
4\ \frac{a^2}{b^2}\ \frac{\partial\ln D}{\partial a}\,.
\end{equation}
Thus, the equation for $a(t)$ is from Eqs.\ (\ref{11}),
(\ref{23}) and (\ref{25})

\begin{equation}\label{26}
\left(\frac{d^2}{dt^2}+8\mu^2\frac{a^2}{b^2}\right)
\frac{\partial\ln D}{\partial a}\ =\ 0\,.
\end{equation}
The expansion of Eq.\ (\ref{26}) is straightforward, though
tedious.

In order to check the validity of Eq.\ (\ref{8}) a simple
system obeying Langevin dynamics was studied numerically.
The system was chosen to be so simple that the integral of
Eq.\ (\ref{8}) may be evaluated "exactly", thus avoiding the
noise of Monte Carlo simulations.
No attempts were made to verify that Eq.\ (\ref{26}) gives the
correct $a(t)$.
Instead, the following procedure was followed.
At a number of fixed $a$-values $b(a)$ was found from the
requirement that the average joint potential energy is
$2\left<U\right>$.
Then the mean-square displacement was evaluated for each $a$
from Eq.\ (\ref{11}) and also as function of time from the
dynamical simulations, allowing an identification of the times
corresponding to the fixed $a$-values.
Finally, the time auto-correlation function was calculated
from Eq.\ (\ref{8}) at the fixed $a$-values.
Figure 2 shows the results for the following time
auto-correlation function $\left<X^3(0)X^3(t)\right>$ for a
particle in a double well potential obeying Langevin dynamics
\cite{mor92}. 
The full curve is the exact time auto-correlation function
found by solving the Smoluchowski equation \cite{van81} and
the dots give the prediction of Eq.\ (\ref{8}).
Results are shown for the cases $\beta=2$ and $\beta=8$ in
dimensionless units.
There is good agreement between prediction and simulation.

To conclude, in this paper a statistical mechanical
approximation for the calculation of time auto-correlation
functions has been proposed.
The formalism assumes a knowledge of the mean-square
displacement in configuration space as a function of time, the
mean-square displacement acting as a "molecular clock".
The remaining "spatial" auto-correlation is a "double"
canonical average (Eq.\ (\ref{8})).
Note that the "double partition function", $D(a,b)$, contains
the ordinary partition function (of configuration space) as a
special case: $Z^2(\beta)=D(0,\beta)$.

The approximation proposed is only useful if
${\left<\Delta X^2(t)\right>}$ is known.
Experimentally, this quantity is accessible via the
intermediate incoherent scattering function.
In some cases phenomenological estimates of the mean-square
displacement as function of time may be given.
Thus, for hopping in a rugged energy landscape where all
minima are equal, the mean-square displacement is
{\bf universal} at low temperatures (except for trivial
scalings), i.e., independent of the barrier height
probability distribution.
This has been shown \cite{dyr94} by the effective medium
approximation and by computer simulations of the 
frequency-dependent conductivity (which is simply related to the
mean-square displacement \cite{sch73}).
Finally, there is the possibility that the mean-square
displacement may be found approximately from Eq.\ (\ref{26})
if Langevin dynamics is assumed, as is common, e.g., in
polymer dynamics \cite{doi86}.

Equation (\ref{26}) signals that Langevin dynamics plays a
special role in the proposed scheme for calculation of time
auto-correlation functions.
A question of considerably interest is if an when Langevin
dynamics can be expected to give the same time 
auto-correlation functions as Newtonian dynamics \cite{low91}.
If the ansatz proposed above is correct, two different
dynamics give the same time auto-correlation functions for any
quantity defined in configuration space, if the two dynamics just
give the same mean-square displacement as function of time.
In this way the ansatz provides a mechanism for the
consistency of any two types of dynamics.

\acknowledgements
 This work was supported by the Danish Natural Science Research
Council.

\begin{figure}
\caption[lgt].
There are two figures.  
Figures and figure caption are available from author by snail
mail, please write an E-mail to order (dyre@ruc.dk).
\end{figure}


\begin{references}
\bibitem{note} The paper was submitted to Phys. Rev. B in the
spring of 1994, but not accepted without revision.
I have not yet taken the time to correct the paper according to
the reviewers suggestions.
One point made in the referee report should be mentioned here:
The reviewer pointed out that Haan in 1979 proposed a related
idea [S. W. Haan, Phys. Rev. A 20, 2516 (1979)].
Haan worked in real 3d-space and not in configuration space, but
his work (of which I was unaware) is clearly built on the same
thinking as presented here.  
The reviewer is thanked for pointing this out.
\bibitem{zwa65} R. Zwanzig, Ann. Rev. Phys. Chem. {\bf 16}, 67
(1965).
\bibitem{boo80} J. P. Boon and S. Yip, {\it Molecular
Hydrodynamics} (McGraw-Hill, New York, 1980).
\bibitem{han86} J.-P. Hansen and I. R. Macdonald, {\it Theory of
Simple Liquids}, 2nd Ed. (Academic Press, New York, 1986).
\bibitem{all87} M. P. Allen and D. J. Tildesley, {\it Computer
Simulations of Liquids} (Clarendon Press, Oxford, 1987).
\bibitem{cic87} {\it Simulations of Liquids and Solids}, Eds. G.
Ciccotti, D. Frenkel, and I. R. Macdonald (North-Holland,
Amsterdam, 1987).
\bibitem{all93} {\it Computer Simulations in Chemical Physics},
Eds. M. P. Allen and D. J. Tildesley (Kluwer Academic Publishers,
Dordrecht, 1993).
\bibitem{bir87} R. B. Bird, R. C. Armstrong, and O. Hassager,
{\it Dynamics of Polymeric Liquids}, 2nd Ed. (Wiley, New York,
1987), Vol. 1. 
\bibitem{chr94} T. Christensen and N. B. Olsen, Phys. Rev. B {\bf
49}, 15396 (1994).
\bibitem{wil75} G. Williams, in {\it Dielectric and Related
Molecular Processes, Specialist Periodical Report, Vol. 2}, Ed.
M. Davies (Chem. Soc., London, 1975), p. 151.
\bibitem{chr85} T. Christensen, J. Physique (Paris) Colloq. {\bf
46}, C8-635 (1985).
\bibitem{bir85} N. O. Birge and S. R. Nagel, Phys. Rev. Lett {\bf
54}, 2674 (1985).
\bibitem{boh93} R. B{\"o}hmer, K. L. Ngai, C. A. Angell, and D.
J. Plazek, J. Phys. Chem. {\bf 99}, 4201 (1993).
\bibitem{mor92} M. Morillo and J. Gomez-Ordonez, Phys. Rev. A
{\bf 46}, 6738 (1992).
\bibitem{van81} N. G. van Kampen, {\it Stochastic Processes in
Physics and Chemistry} (North-Holland, Amsterdam, 1981).
\bibitem{dyr94} J. C. Dyre, Phys. Rev. B {\bf 49}, 11709 (1994).
\bibitem{sch73} H. Scher and M. Lax, Phys. Rev. B {\bf 7}, 4491
(1973).
\bibitem{doi86} M. Doi and S. F. Edwards, {\it The Theory of
Polymer Dynamics} (Clarendon Press, Oxford, 1986).
\bibitem{low91} H. L{\"o}wen, J.-P. Hansen, and J.-N. Roux, Phys.
Rev. A {\bf 44}, 1169 (1991).
\end{references}
\end{document}